\documentstyle[aps]{revtex}
\begin{document}
\title{ Properties of the instantaneous Ergo Surface of a Kerr Black Hole }
\author{ Nicos Pelavas, Nicholas Neary  and Kayll Lake\thanks{Electronic mail: lake@astro.queens.ca}}
\address{Department of Physics, Queen's University, Kingston, Ontario, Canada, K7L 3N6 }   
\date{\today}  
\maketitle
\begin{abstract}
This paper explores properties of the instantaneous ergo surface of a Kerr black hole. The surface area is evaluated in closed form. In terms of the mass ($m$) and angular velocity ($a$), to second order in $a$, the area of the ergo surface is given by $16 \pi m^2 + 4 \pi a^2$ (compared to the familiar $16 \pi m^2 - 4 \pi a^2$ for the event horizon). Whereas the  total curvature  of the instantaneous event horizon is $4 \pi$, on the ergo surface it ranges from $4 \pi$ (for $a=0$) to $0$ (for $a=m$) due to conical singularities on the axis ($\theta=0,\pi$) of deficit angle $2 \pi (1-\sqrt{1-(a/m)^2})$. A careful application of the Gauss-Bonnet theorem shows that the ergo surface remains topologically spherical. Isometric embeddings of the ergo surface in Euclidean 3-space are defined for $0 \leq a/m \leq 1$ (compared to $0 \leq a/m \leq \sqrt{3}/2$ for the horizon).
\end{abstract}


\section{Introduction}

Axially symmetric uncharged black holes are described by the Kerr spacetime \cite{frolov}, one of the cornerstones in our understanding of Einstein's theory of the gravitational field. The Kerr metric \cite{dinverno} is perhaps most familiar in Boyer-Lindquist coordinates $(r,\theta,\phi,t)$
 \begin{eqnarray}
ds^2= \frac{(r^2+a^2cos^2\theta)}{(r^2-2mr+a^2)}dr^2+(r^2+a^2cos^2\theta)d\theta^2 \label{Boyer-Lindquist}
+sin^2\theta(r^2+a^2+ \\ \frac{2mra^2sin^2\theta}{(r^2+a^2cos^2\theta)})d\phi^2
-\frac{4mar sin^2\theta } {(r^2+a^2cos^2\theta)}d\phi dt-(1-\frac{2mr}{(r^2+a^2cos^2\theta)})dt^2, \nonumber
\end{eqnarray} 
or in ``ingoing Eddington-Finkelstein" coordinates $(r,\theta,\overline{\phi},v)$
\begin{eqnarray}
ds^2= 2 dr dv -2 a sin^2\theta dr d\overline{\phi} -\frac{4marsin^2\theta}{(r^2+a^2 cos^2\theta)}d\overline{\phi}dv\label{Eddington-Finkelstein}
+sin^2\theta(r^2+a^2+ \\ \frac{2mra^2sin^2\theta}{(r^2+a^2cos^2\theta)})d\overline{\phi}^2
+(r^2+a^2cos^2\theta)d\theta^2-(1-\frac{2mr}{(r^2+a^2cos^2\theta)})dv^2. \nonumber
\end{eqnarray} 
For black holes $0 \leq a \leq m$. The ergo surface\cite{hawking}, given by the largest root of
\begin{equation}
r^2-2mr+a^2cos^2\theta=0,\label{ergo surface}
\end{equation}
signifies the dominance of the dragging of inertial frames. Static (non rotating stationary) observers are forced to corotate with the black hole within this surface (hence the name \textit{static} or \textit{stationary} limit surface). The surface is second in importance only to the event horizon, given by the largest root of
\begin{equation}
r^2-2mr+a^2=0. \label{horizon}
\end{equation}
The region between the ergo surface and the event horizon is called the \textit{ergo region} or \textit{ergo sphere} and is the site of the Penrose process\cite{mtw}. 
This paper explores intrinsic properties of the instantaneous ergo surface. It was motivated by the fact that whereas the area of the event horizon plays a central role in the laws of black hole mechanics\cite{wald}, it would appear that the area of the ergo surface has not been evaluated in closed form\cite{pandz}. This area is of potential importance for detailed studies of the Penrose process. It is instructive to include with the calculations on the ergo surface, calculations on the event horizon as they are both simpler and rather more familiar. The ergo surface develops conical singularities on the axis so that a study of its topology requires a rather careful application of the Gauss-Bonnet theorem.

\section{Surface Area}

\subsection{Calculation}

A closed two-surface $\Sigma$ with intrinsic coordinates ${x^A}$ and metric $\sigma_{AB}$ has an area defined by
\begin{equation}
A_{\Sigma} = \int \int_{\Sigma}\sqrt{\sigma} dx^{A} dx^{B} \label{area}
\end{equation} 
where $\sigma$ is the determinant of $\sigma_{AB}$ and integration is over the range in $x^A$. For the horizon ($H$), since $r=const$ from (\ref{horizon}), the instantaneous intrinsic metric is found by setting $dr=0$ and $dt=0$ in (\ref{Boyer-Lindquist}) (or equivalently $dr=0$ and $dv=0$ in (\ref{Eddington-Finkelstein})). The result is
\begin{equation}
ds_{H}^2= m^2(x^2+A^2cos^2\theta)d\theta^2+m^2sin^2\theta(2x+\frac{2xA^2sin^2\theta}{x^2+A^2cos^2\theta})d\phi^2 \label{horizonmetric}
\end{equation}
where 
\begin{equation}
x=1+\sqrt{1-A^2},\label{horizonx}
\end{equation}
$x=r/m$, $A=a/m \in [0,1]$, $\theta \in [0,\pi]$ and $\phi \in [0,2\pi)$. The integration is elementary and yields the familiar result
\begin{equation}
A_{H}= 8\pi  m^2(1+\sqrt{1-A^2})\label{horizonarea}.
\end{equation}
For the ergo surface ($S$), it follows from (\ref{ergo surface}) that $r\not=const$. The instantaneous intrinsic metric is found by using  (\ref{ergo surface}) and setting $dt=0$ in (\ref{Boyer-Lindquist}) (or equivalently $dv=0$ in (\ref{Eddington-Finkelstein})).\footnote{The paper\cite{pandz} errs in setting $r=const$ in (\ref{Boyer-Lindquist}) (or equivalently in (\ref{Eddington-Finkelstein})) and then using (\ref{ergo surface}). As a result, the calculation of $A_{S}$ in \cite{pandz} is in error at order $A^2$. Also, it is essential to realize that the coordinates are \textit{not} spherical polars.} The result is
\begin{equation}
ds_{S}^2= \frac{2m^2x(\theta)}{(x(\theta)-1)^2}d\theta^2+2m^2sin^2\theta(x(\theta)+A^2sin^2\theta)d\phi^2 \label{ergometric} 
\end{equation}
where 
\begin{equation}
x(\theta)=1+\sqrt{1-A^2cos^2\theta}, \label{ergox}
\end{equation}
and again $x=r/m$, $A=a/m \in [0,1]$, $\theta \in [0,\pi]$ and $\phi \in [0,2\pi)$. From (\ref{area}) and (\ref{ergometric}) the integration now yields
\begin{eqnarray}
A_{S}= \frac{16 \pi m^2}{A}(\frac{A}{3}+(A^2+2)^{1/4}(E(sin\psi,k)+\\ \label{ergoarea}
(\frac{\sqrt{A^2+2}}{3}- \frac{1}{2})F(sin\psi,k)-\frac{sin\psi \sqrt{1-k^2sin^2\psi}}{(1+cos\psi)}))  \nonumber,
\end{eqnarray}
where $F$ and $E$ are the incomplete elliptic integrals of the first and second kinds respectively\cite{byrd},
\begin{equation}
k=\frac{1}{2}\sqrt{\frac{3+2\sqrt{A^2+2}}{\sqrt{A^2+2}}},
\end{equation} 
and
\begin{equation}
\psi=arccos(\frac{\sqrt{A^2+2}-1+\sqrt{1-A^2}}{\sqrt{A^2+2}+1-\sqrt{1-A^2}}).
\end{equation}

\subsection{Approximations}

 From (\ref{horizonarea}) and (11) we obtain the following approximations:
\begin{equation}
A_{H}= 16\pi m^2-4\pi m^2A^2-\pi m^2A^4-\frac{1}{2}\pi m^2 A^6-\frac{5}{16}\pi m^2 A^8+O(A^{10}),\label{horizonareaapprox}
\end{equation}
and
\begin{equation}
A_{S}= 16\pi m^2+4\pi m^2A^2+\frac{3}{5}\pi m^2A^4+\frac{33}{70}\pi m^2 A^6+\frac{191}{720}\pi m^2 A^8+O(A^{10}).\label{ergoareaapprox}
\end{equation}
The exact forms for $A_{H}$ and $A_{S}$ are compared to their approximations to order $A^2$ and to order $A^8$ in FIGURE 1.

\section{Intrinsic Geometry}

\subsection{Ricci scalar (Gauss curvature)}

Since we are considering two-dimensional cross-sections of the event horizon and ergo surface, the associated Ricci scalar ($R$) (twice the Gauss curvature) is of central importance. The most important feature of this scalar is the fact that it fails to be positive for all $A$. Both on the event horizon and ergo surface the Ricci scalar is minimal (for $A>0$) at the axis ($\theta=0,\pi$). The variation of this minimum with $A$ is shown in FIGURE 2. The Ricci scalar is positive over the event horizon for $A < \sqrt{3}/2 \sim 0.866$ and over the ergo surface for $A < \sim 0.525$. On the horizon the Ricci scalar evaluates to
\begin{equation}
R_{H}=\frac{4x(x^2-3A^2cos^2\theta)}{(x^2+A^2cos^2\theta)^3m^2} ,\label{horizonricci}
\end{equation}
with $x$ given by (\ref{horizonx}). The variation of $R_{H}$ with $\theta$ and $A$ is shown in FIGURE 3. On the ergo surface the Ricci scalar $R_{S}$ shows the same qualitative variation with $\theta$ and $A$ as $R_{H}$. (The variation of $R_{S}$ with $\theta$ and $A$ has been studied previously by Kokkotas\cite{kokkotas}.) $R_{S}$ is shown in FIGURE 4. On the ergo surface the Ricci scalar is given by
\begin{eqnarray} \nonumber 
R_{S}=(-12 A^8  cos^4\theta-16-16 S(\theta)-144 S(\theta) cos^4\theta A^4+104 S(\theta)cos^2\theta A^2 \\ \nonumber
-62 S(\theta) A^6  cos^4\theta + 57 S(\theta) cos^6\theta  A^6+ 86 S(\theta) A^4  cos^2\theta+ 
17 S(\theta) A^6  cos^2\theta \\ \nonumber - 194 A^4  cos^4\theta  + 100 A^4  cos^2\theta  - 28 A^2+ 112 A^2  cos^2\theta  + 22 A^2  cos^2\theta - 20 cos^8\theta  A^8\\ \nonumber
+ 32 A^8  cos^6\theta  - 28 S(\theta) A^2  - 10 S(\theta) A^4+ 118 cos^6\theta  A^6  - 104 A^6  cos^4\theta  - 10 A^4 )/\\ \nonumber
(4m^2  (-8 - 8 S(\theta) A^2  + S(\theta) A^6  cos^2\theta+ 10 S(\theta) A^4  cos^2\theta +  S(\theta) cos^6\theta  A^6 \\ \nonumber
- 2 S(\theta) A^6 cos^4\theta  + 16 S(\theta) cos^2\theta  A^2- 9 S(\theta) cos^4\theta  A^4 -8 S(\theta) - 16 A^4  cos^4\theta - 8 A^2 \\ \nonumber
  +20 A^2  cos^2\theta  + 14 A^4  cos^2\theta + 2 A^6  cos^2\theta  -6 A^6 cos^4\theta  + 4 cos^6\theta  A^6- 2 A^4  - 2S(\theta) A^4 ))  \nonumber 
\end{eqnarray}
where 
\begin{eqnarray} \nonumber
S(\theta)\equiv\sqrt{1-A^2cos^2\theta}.
\end{eqnarray}
Despite its superficial complexity, $R_{S}$ can be evaluated essentially instantaneously ($< 1$ second on any contemporary PC).

\subsection{Total Curvature}

The ``total" curvature of $\Sigma$ is defined by
\begin{equation}
T_{\Sigma} \equiv \frac{1}{2} \int \int_{\Sigma}R\sqrt{\sigma} dx^{A} dx^{B}. \label{total}
\end{equation}
A familiar, but elementary (and for our purposes inadequate) form of the Gauss-Bonnet theorem states that if $\Sigma$ is a compact orientable 2-manifold then
\begin{equation}
T_{\Sigma}=
2\pi\chi(\Sigma), \label{GaussBonnetelm}
\end{equation}
where $\chi(\Sigma)$ is the Euler characteristic of $\Sigma$ (a ``topological invariant" = $2-2n$ where $n$ is the genus of $\Sigma$ equal to the number of ``holes"). On the horizon the calculation of the total curvature is elementary and gives $T_{H} = 4 \pi$, demonstrating the spherical topology of the horizon\cite{smarr}. (The proof is given in \cite{hawking}.) In contrast, we find \footnote{We do not have an analytical proof of (\ref{ergocurvature}), only a precise numerical demonstration with errors less than one part in $10^{18}$.  The errors can be made 
arbitrarily small with increased computational time. This integral has been evaluated previously by Kokkotas\cite{kokkotas}.}
\begin{equation}
T_{S} = 4\pi \sqrt{1-A^2}.\label{ergocurvature}
\end{equation}
$T_{S}$ is shown in FIGURE 5 for convenience. Note that $T_{S}$ varies from $4 \pi$ (for $A=0$) to $0$ (for $A=1$). Clearly (\ref{GaussBonnetelm}) does not hold on the instantaneous ergo surface. This is explained in the next section.

\subsection{Conical Singularities}

\subsubsection{Background}

A right circular cone, of deficit angle $\delta$, can be described intrinsically by
\begin{equation}
ds^2 = d \theta^2+\theta^2 d \phi^2 \label{plane}
\end{equation}
where $\theta \in (0,\infty)$, $\phi \in[0, 2\pi-\delta]$, or equivalently by
\begin{equation}
ds^2 = d \theta^2+(1-\frac{\delta}{2 \pi})^2\theta^2 d \phi^2 \label{planecone}
\end{equation}
where $\phi \in [0,2\pi)$. Within an enveloping Euclidean 3-space, the cone subtends a vertex half angle of $\alpha$ with $sin (\alpha) = 1-\delta/2 \pi$.

Expanding (\ref{horizonmetric}) with (\ref{horizonx}) to order $\theta^2$ about $\theta=0$ (or, equivalently about $\theta=\pi$) up to an ignorable constant factor we obtain (\ref{plane}) with $\phi \in [0,2\pi)$. In contrast, expanding (\ref{ergometric}) with (\ref{ergox}) to order $\theta^2$ about $\theta=0$ (or, equivalently about $\theta=\pi$) up to an ignorable constant factor we obtain (\ref{planecone}) with
\begin{equation}
\delta = 2 \pi (1-\sqrt{1-A^2}).
\end{equation}
That is, the instantaneous ergo surface has conical singularities with $sin(\alpha) = \sqrt{1-A^2}$ 
at the axis ($\theta=0,\pi$). These singularities invalidate the use of (\ref{GaussBonnetelm}).
\medskip

We now give a version of the Gauss-Bonnet theorem which can be used for surfaces with conical singularities.
   
\subsubsection{Generalized Gauss-Bonnet theorem.}

We omit certain conditions (even though they are implicitly assumed) and refer the reader to a standard text (e.g. Do Carmo\cite{docarmo}) for background details. 
The global Gauss-Bonnet theorem assumes that $r \subset M$ is a regular region of an oriented surface $M$.
Let $C_{i},i=1, \ldots, n$ be the closed, simple, piecewise regular curves which form the boundary of r.
Suppose that each $C_{i}$ is positively oriented, and let $\theta_{l},l=1, \ldots, p$ be the set of all external 
angles of the curves $C_{i}$. It is well known that (e.g. \cite{docarmo} pg. 274)

\begin{equation}
\frac{1}{2}\int\int_{r}R \sqrt{\sigma}dx^Adx^B + \sum_{i=1}^{n}\int_{C_{i}}\kappa_{g}(s)ds + \sum_{l=1}^{p}\theta_{l} = 2\pi\chi(r),
\end{equation}
where $\kappa_{g}(s)$ denotes the geodesic curvature with respect to arc length $s$ of $C_{i}$, the integral 
over $C_{i}$ means the sum of integrals in every regular arc of $C_{i}$ and $\chi(r)$ is the Euler-Poincare
characteristic of the region r.
In dealing with non-regular surfaces, such as ones with conical singularities, applying the above form of the Gauss-Bonnet 
theorem yields (incorrect) non-integer values for the Euler characteristic.
\medskip

We reconsider a standard proof of the global Gauss-Bonnet theorem (e.g. as stated in \cite{docarmo} pg. 275) and make the appropriate 
modifications so as to include surfaces with conical singularities.  It suffices to consider only the case where 
the conical singularities are on the interior of the surface, as the conical singularities on the boundary 
are accounted for by the term $\sum_{l=1}^{p}\theta_{l}$ in the global Gauss-Bonnet theorem.
\medskip

Let $r$ be a region of an oriented surface $M$ with conical singularities on the interior and denote by $V_{ik}$ the 
number of conical singularities in $r$.  Consider a triangulation $t$ of $r$ such that the conical singularities 
of $r$ are a subset of the interior vertices of the triangulation.  Given a triangle $t_{j}$ of $t$ there
are no conical singularities in its interior, so the local Gauss-Bonnet theorem applies to each triangle of the 
triangulation.  Now, if $V_{i}$ denotes the number of internal vertices of the triangulation, 
and $V_{if}$ denotes the number of internal vertices of the triangulation that are not conical singularities then 
$V_{i}=V_{if}+V_{ik}$.  The essential point here is that the sum of the face angles about each vertex of $V_{if}$ 
is $2\pi$ whereas the sum of the face angles about each conical singularity is $\phi_{s}\neq 2\pi$ where 
$s=1,...,V_{ik}$.  Applying to every triangle the local Gauss-Bonnet theorem, adding up the results and 
simplifying, we have

\begin{equation}
\frac{1}{2}\int\int_{r}R \sqrt{\sigma}dx^Adx^B + \sum_{i=1}^{n}\int_{C_{i}}\kappa_{g}(s)ds + \sum_{l=1}^{p}\theta_{l} + \sum_{s=1}^{V_{ik}}\delta_{s} = 2\pi\chi(r) \label{coneGB}
\end{equation}
where $\delta_{s}=2\pi - \phi_{s}$ is the deficit angle of a conical singularity. (This is actually a discreet 
measure of the curvature at the conical singularity.)  As a special case consider a closed polyhedron $P$. Then (\ref{coneGB}) simplifies to
\begin{equation}
\sum_{s=1}^{V_{ik}}\delta_{s} = 2\pi\chi(P)
\end{equation}
where $V_{ik}$ denotes the number of vertices of $P$ and $\delta_{s}=2\pi - \phi_{s}$, $\phi_{s}$ is the sum of the face angles around a vertex. This is exactly Descartes result for polyhedral curvature (Descartes curvature).

\medskip

In summary, the ``new" term in (\ref{coneGB}) is a re-expression of the expected deficit angle with a triangulation in which the singularity is located at a vertex.

\medskip

We now illustrate the use of this theorem with a cone, a rugby ball and finally the instantaneous ergo surface. These applications make use of isometric embeddings in Euclidean 3-space ($E^3$).

\subsubsection{Cone}

Consider the cone with an $E^{3}$ parametrization given by
\begin{equation}
x(\rho,\phi)=(\rho\sin(\alpha)\cos(\phi),\rho\sin(\alpha)\sin(\phi),-\rho\cos(\alpha))
\end{equation}
where $\rho \in (0, \infty)$, $\phi \in [0, 2\pi)$, and $\alpha$ is the angle from the z-axis to the surface of the cone.
This cone has the intrinsic metric
\begin{equation}
ds^2=d\rho^2+\rho^2\sin(\alpha)^2d\phi^2.
\end{equation}
Consider the closed unit speed curve $C$ given by
\begin{equation}
x^{a}(s)=\left(\rho_{0},\frac{s}{\rho_{0}\sin(\alpha)}\right)
\end{equation}
where $\rho_{0}$ is a constant, and $C$ is oriented so that it encloses the conical singularity at the origin.
Denote the region enclosed as $\Sigma$. In what follows we calculate the Euler characteristic of this region.
The geodesic curvature is a constant along $C$, and is given by $\kappa_{g}(s)=\frac{1}{\rho_{0}}$.  Since $\Sigma$ has 
one conical singularity $V_{ik}=1$. The total face angle is $\phi_{1}=2\pi\sin(\alpha)$ and so the 
deficit angle is $\delta_{1}=2\pi-2\pi\sin(\alpha)$.  Applying (\ref{coneGB}) to $\Sigma$, and observing that 
the Gaussian curvature for a cone vanishes, we have
\begin{equation}
\int_{C}\kappa_{g}(s)ds + \delta_{1} = 2\pi\chi(\Sigma).
\end{equation}
That is,
\begin{equation}
2\pi\sin(\alpha)+2\pi-2\pi\sin(\alpha) = 2\pi\chi(\Sigma).
\end{equation}
Therefore $\chi(\Sigma)=1$, which is what we expect since $\Sigma$ is homeomorphic to a disk.

\subsubsection{Rugby Ball}

Consider the metric
\begin{equation}
ds^2=\frac{1}{1-A^2}d\phi^2+\sin(\theta)^2d\phi^2
\end{equation}
where $\theta \in (0,\pi)$, $\phi \in[0,2\pi)$, and $A$ is a constant such that $A \in [0,1)$. $A=0$ returns the 
metric for the unit sphere. If $A \in (0,1)$ the closed surface looks like a rugby ball when embedded in Euclidean 3-space.

Denoting the surface by $\Sigma$ 
it has an $E^3$ parametrization
\begin{equation}
x(\theta,\phi)=\left(\sin(\theta)\cos(\phi),\sin(\theta)\sin(\phi),\frac{1}{\sqrt{1-A^2}}E(\cos(\theta),\sqrt{1-A^2})\right).
\end{equation}
$\Sigma$ has two conical singularities, one at $\theta=0$ and the other at $\theta=\pi$ so that $V_{ik}=2$.
It is straight forward to show that the deficit angles are $\delta_{1}=\delta_{2}=2\pi-2\pi\sqrt{1-A^2}$, 
and that the Ricci scalar is given by  $R=2(1-A^2)$.  Applying (\ref{coneGB}) to $\Sigma$ we have
\begin{equation}
T_{\Sigma}+\sum_{s=1}^{2}\delta_{s} = 2\pi\chi(\Sigma),
\end{equation}
so that
\begin{equation}
4\pi\sqrt{1-A^2}+2\pi-2\pi\sqrt{1-A^2}+2\pi-2\pi\sqrt{1-A^2}=2\pi\chi(\Sigma).
\end{equation}
We conclude that $\chi(\Sigma)=2$, which is what we expect, since $\Sigma$ is homeomorphic to a sphere.

\subsubsection{Ergo Surface}

As shown in section $(1)$ above, the ergo surface has two conical singularities, 
at $\theta=0$ and at $\theta=\pi$, so $V_{ik}=2$.  The deficit angles are
 $\delta_{1}=\delta_{2}=2\pi-2\pi\sqrt{1-A^2}$.  Applying (\ref{coneGB}) to $S$ we have
\begin{equation}
T_{S}+\sum_{s=1}^{2}\delta_{s} = 2\pi\chi(S),
\end{equation}
so that from (\ref{ergocurvature})
\begin{equation}
4\pi\sqrt{1-A^2}+ 4\pi-4\pi\sqrt{1-A^2}= 2\pi\chi(S). \label{ergoeuler}
\end{equation}
Therefore $\chi(S)=2$ for $A \in [0,1]$, and the ergo surface is 
homeomorphic to a sphere.  Notice that when $A=1$ the sum of the face angles around the conical singularities is $\phi_{s}=0$ so that cusps have formed but topologically $S$ hasn't changed.

\section{Images}

\subsection {Kerr Coordinates}

Images of the ergo surface in terms of pseudo-Cartesian coordinates can be generated by the definitions
\begin{eqnarray}
X = \sqrt{x^2+A^2}sin(\theta)cos(\phi-arctan(\frac{A}{x}))\\ \nonumber 
Y =  \sqrt{x^2+A^2}sin(\theta)sin(\phi-arctan(\frac{A}{x}))\\ 
Z = x cos(\theta) \nonumber
\end{eqnarray}
where $x$ is given by (\ref{horizonx}) on the horizon and (\ref{ergox}) on the ergo surface. An example is shown in FIGURE 6. Although these images are common in texts (often poorly drawn), it is important to realize that they are \textit{not} embeddings. Accurate plots have been given previously by Sharp\cite{sharp}.

\subsection{Embeddings in $E^3$}

We are interested in isometric embeddings in Euclidean 3-space ($E^3$) of 2-surfaces ($\Gamma$) with intrinsic metrics of the form
\begin{equation}
ds^2_{\Gamma} = g_{\theta \theta}(\theta)d \theta^2 + g_{\phi \phi}(\theta) d \phi^2. \label{2surface}
\end{equation}
Define
\begin{eqnarray}
x = F(\theta) cos(\phi)\\ \nonumber
y = F(\theta) sin(\phi)\\
z = G(\theta) \nonumber
\end{eqnarray}
so that
\begin{equation}
d x^2 + d y^2 + d z^2 = (F^{' 2} + G^{' 2})d \theta^2 + F^2 d \phi^2.
\end{equation}
The surface $\Gamma$ is embedded in $E^3$ via the identifications
\begin{equation}
F^{' 2} + G^{' 2}=g_{\theta \theta} \label{iso1}
\end{equation}
and
\begin{equation}
F^{2} = g_{\phi \phi}. \label{iso2}
\end{equation}
The condition on $\Gamma$ that the embedding exists is

\begin{equation}
g_{\theta \theta} \geq (\frac{d \sqrt{g_{\phi \phi}}}{d \theta})^2. \label{embedding}
\end{equation}
Whereas equality in condition (\ref{embedding}) gives zero  Gaussian curvature,
inequality is \textit{not} equivalent to positive Gaussian curvature.

\medskip

Following the procedure outlined above we obtain the isometric embeddings shown in FIGURE 7.

\medskip
 
Embeddings of the complete ergo surface are defined for $a/m \in [0,1]$. This has been pointed out previously by Sharp \cite{sharp} and Kokkotas\cite{kokkotas}. For the horizon, however, complete embeddings are defined only for $a/m \in [0,\sqrt{3}/2]$. For $a/m > \sqrt{3}/2$ condition (\ref{embedding}) fails near the axis ($\theta=0,\pi$). This failure coincides with (but is \textit{not}\footnote{
This failure of an isometric embedding in $E^3$ is frequently, but erroneously, attributed to negative Gauss curvature. See, for example, the (incorrect) discussion given by Lightman \textit{et al}\cite{lightman}. A theorem by Efimov\cite{efimov} (a generalization of Hilbert's theorem) states that a complete surface with Gauss curvature $K <0$ cannot be isometrically immersed into $E^3$. Neither the ergo surface nor the horizon satisfy the conditions of Efimov's theorem since $K\not<0$ everywhere. }
a consequence of) the onset of negative Gauss curvature (see Figures 3 and 4).

\medskip

Relaxation of the isometric requirements (\ref{iso1}) and (\ref{iso2}), say by a conformal factor $C^2(\theta)$ on the RHS, allows the resultant generalized form of (\ref{embedding}) to be satisfied over the whole horizon for $a/m \in [0,1]$ with a suitable choice for $C^2(\theta)$. An example is shown in FIGURE 8 for which $C^2(\theta)=\frac{1}{2}exp(cos^2(\theta))$.

\section{Summary}

The surface area of the instantaneous ergo surface has been given in closed form and compared with the event horizon.  The former increases monotonically whereas the later decreases monotonically with increasing $a/m$.  In terms of the mass ($m$) and angular velocity ($a$), to second order in $a$, the area of the ergo surface is given by $16 \pi m^2 + 4 \pi a^2$ (compared to the familiar $16 \pi m^2 - 4 \pi a^2$ for the event horizon). The Ricci scalar has be calculated for 
both surfaces and qualitatively it displays similar features. It reaches a minimum at the poles and a maximum at the equator. However, whereas the total curvature for the event horizon is $4\pi$, on the ergo surface it ranges from 
$4\pi$ (for $a=0$) to $0$ (for $a/m=1$).  This variation results from conical singularities of deficit angle $2 \pi (1-\sqrt{1-(a/m)^2})$ at the poles. A careful application of the Gauss-Bonnet theorem shows that the ergo surface is topologically spherical for $0 \leq a/m \leq 1$. Finally, isometric embeddings of the ergo surface in Euclidean 3-space are defined for $0 \leq a/m \leq 1$ compared to $0 \leq a/m \leq \sqrt{3}/2$ for the horizon. Conformal embeddings of the horizon can be defined for $0 \leq a/m \leq 1$.

\section{Acknowledgments}

This work was supported in part by an OGSST scholarship to NP and by an NSERC grant to KL. Portions of this work were made possible by use of \textit{GRTensorII}\cite{grt}. It is a pleasure to thank Eric Woolgar for helpful discussions and Kostas Kokkotas for pointing out some references we had not found. The comments of a referee helped to improve the presentation of this paper.

\medskip
\noindent {\bf Figure Captions}
\medskip

{\bf Figure 1} $A_{H}$ and $A_{S}$ as functions of $A (\equiv \frac{a}{m})$. The upper curve is for the ergo surface, and the lower curve is for the event horizon. The approximations to order $A^2$ and to order $A^8$ are shown. The accurate curves have the greatest vertical range.

{\bf Figure 2} The Ricci scalar on the axis ($\theta=0,\pi$) as functions of A. The upper curve is for the event horizon, 
and the lower curve is for the ergo surface.

{\bf Figure 3} Variation of the Ricci scalar over the event horizon. The values of $A$ (read top to bottom on the RHS) are $(1,0.995,0.99,0.98,0.96,0.94,0.92,0.9,0.8,0.7,0.6,0.5,0.4,0.3,0.2,0.1,0)$.

{\bf Figure 4} Variation of the Ricci scalar over the ergo surface. The values of $A$ (read bottom to top on the LHS) are $(1,0.9,0.8,0.7,0.6,0.5,0.4,0.3,0.2,0.1,0)$.

{\bf Figure 5} Variation of the total curvature ($T_{S}/2 \pi$) over the ergo surface as a function of $A$.

{\bf Figure 6} Pseudo Cartesian images of the ergo surface and event horizon. For these images $A=1$. A cut to show the horizon is shown. These images are \textit{not} embeddings in Euclidean space.

{\bf Figure 7} Isometric embeddings of the ergo surface and event horizon. For these images $A=0.9$. A cut to show the (partial) horizon, which is not isometrically embeddable around the poles, is shown. The polar radius of the ergo surface diverges for $A=1$\cite{kokkotas}\cite{sharp}. These images \textit{are} embeddings in Euclidean space.

{\bf Figure 8} Isometric embedding of the ergo surface but conformal embedding of the horizon ($A=0.9$). The conformal factor on the horizon is $\frac{1}{2}exp(cos(\theta)^2)$.
 \end{document}